\magnification=1200

\vskip 2.5in

\font\bigbf=cmbx9  scaled\magstep2 \vskip 0.6in \centerline{\bigbf
 Lorentzian Wormholes }
 \vskip 0.2in
\centerline{\bigbf in the Friedman-Robertson-Walker  Universe}

\vskip 0.4in \font\bigtenrm=cmr10 scaled\magstep1
\centerline{\bigtenrm  B. Mirza$  $, M. Eshaghi$  $ , S.
Dehdashti$  $  }

\vskip 0.2in

\centerline{\sl Department of Physics, Isfahan University of
Technology, Isfahan 84154, Iran }



\vskip 0.6in \centerline{\bf ABSTRACT} \vskip 0.1in

The metric of some Lorentzian wormholes in the background of the
FRW universe is obtained. It is shown that for a de Sitter
space-time the new solution is supported by Phantom Energy. The
wave equation for a scalar field in such backgrounds is
separable. The form of the potential for the Schr\"{o}dinger type
one dimensional wave equation is found.



\vfill\eject

\vskip 2in

\noindent{1. \bf   Introduction } \vskip 0.1in

\noindent Since the pioneering article by Morris and Thorne [1],
Lorentzian wormholes have attracted a lot of attention. A number
of static and dynamic wormholes have been found both in general
relativity and in alternative theories of gravitation. There are
several reasons that support the interest in these solutions. One
of them is the possibility of constructing time machines [2,3].
Another reason is related to the nature of the matter that
generate these solutions. One major problem with the existance of
traversable wormholes is that the matter required to support such
a geometry essentially violates the energy conditions for general
relativity [4,5,6,7]. To get around weak energy condition
violations, there have been some works in non-standard gravity
theories. Brans-Dicke gravity supports static wormholes both in
vacuum [8] and with matter content that do not violate the weak
energy condition by itself [9]. There are analysis in several
others alternative gravitational theories, such as
Einstein-Cartan models [10], Einstein-Gauss-Bonnet [11] and $R+
R^2$ [12]. In all cases the violation of weak energy conditionis
a necessary condition for static wormhole to exist, although
these changes in the gravitational action allows one to have
normal matter while relagating the exoticity to non-standard
fields. This leads to explore the issue of weak energy condition
violation in non-static situations. Time dependence allow one to
move the energy condition violations around in time
[13,14,15,16,17,18], However for some comments see [19].
(Unfortunately radial null geodesics through the wormhole will
still encounter energy condition violations, subject to suitable
technical qualifications.)

In this paper we have obtained evolving worm holes in the FRW
universe with a special choice for their shape function and with
a Phantom Energy. For other ideas about growing wormholes by
Phantom Energy accretion see [20, and refs]. The paper is
organised as follows: In Section 2, the metric of a traversable
wormhole in the FRW universe is obtained and then generalised to
arbitrary dimensions. Finally the metric in an asymptotically de
Sitte space is given and it is shown that the energy density is
positive but the weak energy condition is violated. In the last
section a one dimensional Schr\"{o}dinger like equation is
obtained which corresponds to the scalar wave equations in this
kind of geometry.

 \vskip 0.3in \noindent {2. \bf   \ A traversable wormhole in the FRW universe} \vskip 0.1in

\noindent In this part we are going to apply a new method [21]
for obtaining a new solution for the evolving wornhole. We
consider the following metric for a spherically symmetric and
static wormhole

$$ ds^2=-e^{-2 \phi(r) } dt^2+ {dr^2\over (1- {b(r)\over r})}+ r^2
(d \theta^2 +sin^2 {\theta} \ d\phi^2) \eqno(1) $$

\noindent where $\phi(r)$ is the redshift functionand $b(r)$ is
the shape function. Here we shall restrict ourselves to the case
$ \phi(r)=0 $ and $b(r)=b_0^m/r^m $, where $b_0$ is the radius of
the throat of the wormhole and m is a real positive number. The
following transformations can be use to rewrite  the metric (1)
in the isotropic spherical coordinates.

$$ s=2^{-2/ m}\ell, \ \ \ \ t=2^{-2/m}v, \ \ \ \ {b_0^m\over 2}={b^m},
 \ \ \ \  r={x\over 2^{1/m}}(1+{b^m\over x^m})^{2/ m}  \eqno(2) $$

\noindent So we get to

$$ d\ell^2 = -d v^2 + (1+{b^m\over x^m})^{4/ m}(dx^2+ x^2 d \theta^2 +
x^2  sin^2 {\theta} \ d\phi^2) \eqno(3) $$

\noindent The metric for the FRW universe is given by

$$ d\ell^2= -dv^2 + {a^2(v)\over{(1+k x^2/4)^2}}(dx^2+ x^2 d \theta^2 +
x^2  sin^2 {\theta} \ d\phi^2) \eqno(4)$$

\noindent where $a(v) $ is the scale factor and k gives the
curvature of the universe. By comparing $(3)$ and $(4)$ we set
the metric for a wormhole embedded in the FRW universe as follows

$$ d\ell^2 = -A^2(v,x)dv^2+ B^2(v,x) (dx^2+ x^2 d \theta^2 +
x^2  sin^2 {\theta} \ d\phi^2) \eqno(5) $$

\noindent By considering  $G_{01} =0$ one has

$$A(v,x)=f(v){\dot B \over 2 B} \eqno(6)$$

\noindent where dot denotes the derivative with respect to $v$
and $f(v)$ is an arbitrary function of $v$. By comparing the
$g_{11}$ terms in (3) and (5), the possible form for the function
$B(x,v)$ is,

$$B(v,x) = [w(v,x)+ {q(v)\over x^m}]^{2/m} \eqno(7) $$

\noindent By inserting  eq. (7) in (6) we arrive at

$$A(v,x)={{f\over m}\ {{(\dot w + {\dot q\over x^m})}\over {(w+{q\over x^m})}}}
\eqno(8)  $$

\noindent In the case of $v=const$ and the asymptotically flat
conditions, $A(v=const,x)$ should be reduced to the $ \sqrt
{-g_{00}} $ term in eq.(3). We infer that the following
identities should always hold

$$ {{f \dot w }}= m w, \ \ \ \ \   \ \ \ \  f \dot q=m q \eqno(9) $$

\noindent  We may define $q$ as

$$ q(v)= b^m d^m(v)  \ \ \ \ \ \ \longrightarrow  \ \ \ \  f={d\over\dot d} \eqno(10) $$

\noindent where $d(v)$ is an arbitrary function which is related
to the scale factor of the universe. For the two limiting cases
of small and large $x$ values eq. (5) reduces to (3) and (4)
respectively, and so we can identify the form of $w(v,x)$ which
is similar to the exact solution which has already been  obtained
for Black holes. [21]

$$ w(v,x) = {{d^m(v)}\over{(1+k x^2/4)^{m/2}} }\eqno(11) $$

\noindent The final form of the metric for a traversable wormhole
in the background of of FRW universe is given by

$$ d \ell^2= -d v^2+ a^2(v)[{1\over {(1+k x^2/4)^{m/2}}}+ {b^m\over
x^m}]^{4/m}(dx^2+ x^2d^2 \Omega) \eqno(12)
$$

\noindent where $a^2(v)=d^4(v)$ is the scale factor. This
calculation can be  extended  to  $(3+n)-$dimensional wormholes.
After a straightforward calculation for a $(3+n)-$dimensional
traversable  wormhole in the background of FRW universe,  we
arrive at the following form for the metric

$$ d\ell^2=- dv^2 + a^2(v)[{1\over {(1+k x^2/4)^{m n /2}}}+ {b^{mn}\over
x^{mn}}]^{4/{m n}}(dx^2+ x^2 d^2 \Omega_{n+1}) \eqno(13) $$

\noindent  Equation (13) suggest
 a kind of duality for wormholes in different dimensions.
 Consider $ p=m n $, then it is resonable to expect that in a scattering process
 a $(3+n)$-dimensional wormhole with an index m in (13) has a similar behaviour to another wormhole in
 $(3+1)$-dimensions with index $p$. There is also another kind of duality that is implied by (13), a
 $(3+n)$-dimensional wormhole with an index m has a similar behaviour to a $(3+m)$-dimensional wormhole
  with an index n. In the following section we will consider a de Sitter
space and show that for some special choices of the shape function
 the energy density could be positive.


\vskip 0.3in \noindent{3. \bf  Asymptotically  de Sitter
space-time } \vskip 0.1in

\noindent For $a(t)=exp(H t), H=const$ and $ k=0 $, the metric
(12) can be written as

$$ d s^2 = -d t^2+ e^{2 H t} [1+ {{b^m}\over
{r^m}}]^{4/m}(dr^2+ r^2 d^2 \theta + r^2 sin^2 \theta d\phi^2)
\eqno(14)
$$

\noindent For the large values of $r$, (14) reduces to the de
Sitter universe and for small values of x it describes a
traversable wormhole. So we may suggest (14) as the metric for a
traversable wormhole in the de Sitter universe. The Einstein
equation is defined  by

$$ G_{\mu \nu}- \Lambda g_{\mu\nu} =-8 \pi T_{\mu \nu} \eqno(15)$$

\noindent where we assume that in this case $\Lambda=3 H^2$,
Using an orthonormal reference frame with basis

$$ {\bf e}_{\hat t}= {\bf e}_t  $$

$$ {\bf e}_{\hat r}= { {{\bf e}_r }\over e^{H t}\  (1+ {b^m\over r^m})^{2/m}}$$

$$ {\bf e}_{\hat \theta}= { {{\bf e}_\theta }\over e^{H t} r \ (1+ {b^m\over r^m})^{2/m}}$$

$$ {\bf e}_{\hat \phi}= { {{\bf e}_\phi }\over e^{H t} r\  sin\theta \
(1+ {b^m\over r^m})^{2/m}} \eqno(16)  $$

\noindent the energy momentum tensor of the perfect fluid is
written as

$$ T_{\hat \mu \hat \nu}=diag(\rho,p,p,p) \eqno(17) $$

\noindent where $\rho$ denotes the energy density and $p$ is the
pressure in the fluid. One may write the energy momentum tensor
$(17)$ as

$$ T_{\hat \mu \hat \nu}=diag(\rho,-\tau,p,p) \eqno(18)$$

\noindent where $\tau $ denotes the stress (opposite to the
pressure) in the $\bf e_{\hat r}$ direction. It can be shown that
in the orthonormal frame $(16)$, the metric $(14)$ becomes

$$ g_{\hat \mu \hat \nu}=diag(-1,1,1,1)\eqno(19)$$

\noindent After a straightforward calculation we arrive at

$$ \rho ={e^{-2Ht}\over 2 \pi}{ {{b^m}(1-m)}\over{r^{m+2} (1+{b^m\over r^m})^{{4\over m }+2} }}\eqno(20)$$

$$ -\tau ={-e^{-2Ht} \over {2 \pi}} {{{b^m}}\over{r^{m+2} (1+{b^m\over r^m})^{{{4\over m }+2}} }} \eqno(21)$$

\noindent The matter with the property, energy density $\rho
> 0$ but pressure $ p< - \rho< 0$ is known as Phantom Energy. So
if $0 < m < 1$  the energy source for the new solution (14) is
Phantom Energy.


 \vskip 0.3in
\noindent{3. \bf Scalar wave equation} \vskip 0.1in

\noindent  The wave equation of the minimally coupled massless
scalar field in a background which is defined by the metric in
equation (12) is given by

$$ \nabla^{\mu} \nabla_{\mu}\Phi = {1\over {\sqrt {- g}}} \partial_{\mu} ({\sqrt {- g}} g^{\mu \nu}
\partial_{\nu} \Phi ) \eqno(22) $$

\noindent In spherically symmetric space-time, the scalar field
can be separated by variables

$$ \Phi_{lm}= Y_{lm}(\vartheta , \varphi) {u_l (r,t)\over r} \eqno(23)$$

\noindent where $Y_{lm}$ is the spherical harmonics and $l $ is
the quantum angular momentum. If $l=0$ and the scalar field $\Phi
(r)$ depends on r, the wave equation becomes the following
relation

$$ r^2 ({1\over {(1+ {{k r^2}\over 4})^{n\over2}} }+ {b^n\over{r^n}})^{2\over n}\
 {\partial \over {\partial r}}\Phi = const= A \eqno(24)  $$

 \noindent  Thus the static scalar wave without propagation is easily found as the integral form of

$$ \Phi= A \int r^{-2} ({1\over {(1+ {{k r^2}\over 4})^{n\over2}} }+
{b^n\over{r^n}})^{- 2\over n} \ dr \eqno(25) $$

\noindent If the scalar field depends on $r, t;$ the wave equation
after the separation of variables $(\vartheta , \varphi)$ become

$$ - {{{\partial^2}u_l}\over {\partial t_*^2}} + {{{\partial^2}u_l}\over {\partial r_*^2}}=V_l u_l
 \eqno(26) $$

\noindent where the potential is

$$ V(r) = {1\over ({1\over {(1+ {{k r^2}\over 4})^{n\over2}} }+ {b^n\over{r^n}})^{2\over
n}} \ [ {L^2\over r^2} + {2\over n}\ { {{\partial\over \partial r
} ln({1\over {(1+ {{k r^2}\over 4})^{n\over2}} }+
{b^n\over{r^n}}) }\over ({1\over {(1+ {{k r^2}\over
4})^{n\over2}} }+ {b^n\over{r^n}})^{ 2\over n}} ] \eqno(27)
$$

\noindent where $L^2 =l(l+1) $ is the square of the angular
momentum and the proper distance $r_*$ and time $t_*$ have the
following relations to $r$ and $t$,

$${\partial \over {\partial r_*}}= ({1\over {(1+ {{k r^2}\over 4})^{n\over2}} }+ {b^n\over{r^n}})^{ 2\over n}\
{\partial \over {\partial r}}  \eqno(28)$$

$$ {\partial^2 \over {\partial t_*^2} }= a {\partial \over {\partial t} } (a^3 {\partial \over {\partial t} })
\eqno(29)$$

\noindent The properties of the potential can be determined by the
values $n$, $l$, $b$, $k$ and its shape. If the time dependence
of the wave is harmonic as $u_l(r,t_*)={\hat u}_l(r,w_*) e^{-i w_*
t_*}$ the equation becomes

$$[{ d^2\over {dr_*^2}}+w_*^2-V_l (r)]{\hat u}_l(r,w_*)=0 \eqno(30) $$

\noindent It is just the Schr\"{o}dinger equation with energy
$w_*^2$ and potential $V_l(r)$. If $k=0$, $V_l(r)$ approaches
zero as $r\longrightarrow \infty $, which means that the solution
has the form of the plane wave  $ u_l\approx e^{\pm i w_* r_*} $
asymptotically. The result shows that if a scalar wave passes
through the wormhole the solution is changed from $ e^{\pm i w
r_*}$ into $e^{\pm i w_* r_*}$, which means that the potential
affects the wave and experience the scattering. This part may be
continued by using the method which is used in calculation of the
quasinormal modes. (for a comprehensive review, see [22]).

 \vskip 0.3in
\noindent{3. \bf Conclusion} \vskip 0.1in

\noindent In this work the metric for a traversable wormhole in
the FRW universe is obtained. It is shown that for such wormholes
the energy density may be positive although the weak energy
condition is violated. The scalar wave equation in this space
time is reduced to an effective one dimensional Schr\"{o}dinger
equation, which is helpful for investigating the quasi-normal
modes.




\vskip 0.2in \centerline{\bf \ \  References} \vskip 0.1in

\noindent [1] \ M. Morris and K. Thorne, Am. J. Phys. {\bf 56},
395 (1980)

\noindent [2] \ M. Morris, K. Thorne and U. Yurtserver, Phys.
Rev. Lett. {\bf 61}, 1446 (1988).

\noindent [3] \ V. Forlov and I. Novikov, Phys. Rev. {\bf D42},
1913 (1990).

\noindent [4] \ D. Hochberg and M. Visser, Phys Rev. {\bf D56},
4745 (1997).

\noindent [5] \ D. Hochberg and M. Visser, Phys Rev. {\bf
D58},044021 (1998).

\noindent [6] \ D.Ida and S.A. Heyward, Phys Lett {\bf A260}, 175
(1999).

\noindent [7] \ J.L.Friedman, K. Schliech and D.M. Witt, Phys.
Rev. Let. {\bf 71}, 1486 (1993).

\noindent [8] \ A. Agnese and M. La Camera, Phys. Rev. {\bf D51},
2011 (1995).

\noindent [9] \ L. A. Anchordoqui, et al. Phys. Rev. {\bf D55},
5226 (1997).

\noindent [10] \ L. A. Anchordoqui, Mod. Phys. Lett. {\bf A13},
1095 (1998).

\noindent [11] \ B. Bhawal and S. Kar, Phys. Rev. {\bf D46},2464
(1992).

\noindent [12] \ D. Hochberg, Phys. Lett. {\bf B251}, 349 (1990).

\noindent [13] \  T.A. Roman, Phys. Rev. {\bf D47}, 1370 (1993).
gr-qc/9211012.

\noindent [14] \ D. Hochberg and T.W. Kephart, Phys. Rev. Lett.
{\bf 70}, 2665 (1993). gr-qc/9211006.

\noindent [15] \ S. Kar, Phys Rev {\bf D49}, 862 (1994).

\noindent [16] \ S. Kar and D. Sahdev, Phys. Rev. {\bf D53}, 722
(1996).

\noindent [17] \ A. Wang and P. Letelier, Prog. Theor. Phys. {\bf
94}, 137 (1995). gr-qc/9506003.

\noindent [18] \ S. W. Kim, Phys. Rev. {\bf D53}, 6889 (1996).

\noindent [19] \ D. Hochberg and M. Visser, Phys. Rev. Lett. {\bf
81}, 746 (1998).

\noindent [20] \ R.F.  Gonzalez-Diaz, astro-ph/0507714.

\noindent [21] \ C. J. Gao and S. N. Zhang, Phys Lett {\bf B595},
28 (2004), gr-qc/0407045.

\noindent [22] \ K. D. Kokkotas and B.G. Schmidt, " Quasi-Normal
modes of Stars and Black Holes",

\noindent \ \ \ \ \ \ Living Reviews in Relativity:
http://relativity.livingreviews.org/Articles/lrr-

\noindent \ \ \ \ \ \ 1999-2/index.html.

\vfill\eject

\bye